\documentclass[11pt,leqno]{article}   

\usepackage[T1]{fontenc}
\usepackage[utf8]{inputenc}
\usepackage{authblk}
\usepackage[all]{xy}

\usepackage{amsmath,amstext,amsthm,amsfonts,amssymb,color,latexsym}
\numberwithin{equation}{section}

\newcommand{\ic}{\text{i}} 

\usepackage{epsfig}

\newcommand{\beano}{\begin{eqnarray*}}
\newcommand{\enano}{\end{eqnarray*}}
\newcommand{\ena}{\end{eqnarray}}

\newcommand{\be}{\begin{equation}}
\newcommand{\ee}{\end{equation}}
\newcommand{\en}{\end{equation}}

\newcommand{\ba}{\begin{array}}
\newcommand{\ea}{\end{array}}

\renewcommand{\em}{\it}

\newcommand{\bg}{\begin{gathered}}
\newcommand{\eg}{\end{gathered}}

\newcommand{\bea}{\begin{eqnarray}}
\newcommand{\eea}{\end{eqnarray}}

\title{Noncommutative Landau problem and shifted energy levels}
\author[]{Nurisya M. Shah
\thanks{Work supported by UPM-IPM Grant (9473100) and GP-IPS (9645700), Malaysia}}

\affil[]{Department of Physics, Faculty of Science, Universiti Putra Malaysia, 43400 UPM Serdang, Selangor, Malaysia}
\affil[]{Laboratory of Computational Sciences \& Mathematical Physics, Institute for Mathematical Research, Universiti Putra Malaysia, 43400 UPM Serdang, Selangor, Malaysia}

\begin{document}

\maketitle
\begin{abstract}
In this paper we discuss a Landau levels problem within the framework of noncommutative configuration space and phase space. We show that the associated energy levels are being shifted in terms of the noncommutative parameter and can be directly obtained from its associated generalized Landau Hamiltonian. A straight forward way of constructing the shifted energy levels is just via a proper defined Bopp shift applied onto the spatial and momenta operators that are used to describe the Landau levels in symmetric gauge potential. 
\end{abstract}



\section{Introduction}

A concrete physical example of a situation where the introduction of noncommutativity makes a difference in the observed energy spectrum of a quantum mechanical system, is by inspecting the Landau quantization problem.  The Landau levels problem described a charged particle (for e.g. electron) that is placed in a constant magnetic field, perpendicular to the $z$-axis,  under certain gauge field. One can find extensive literature which in particular, have properly mentioned NC Landau problem as their article's title. Results shown are on  the effect of NC feature on the corresponding Landau energy spectrum. This includes selection from noncommutative geometry~\cite{hor}, noncommutative field theory~\cite{sza}, function spaces~\cite{Du:Kang:Wang} and many more. 

For example, noncommutative Landau problem on a plane, a sphere and a torus have also been considered~\cite{nair, mor, gros, geloun}. Relativistic oscillator in the external magnetic field under the noncommutative configuration space, has also become of particular interest, see for e.g.~\cite{man:rai, panella, hou, mirza}. On a detail discussion and analysis of the relationship of the constant and non-constant magnetic field on the NC configuration space are given in~\cite{delduc, Gan:Saha:Gak}. Further realizing NC Landau feature which relate to coherent state, electrodynamics and quantum dots are from~\cite{Houn:Isiaka, diao} and~\cite{pal:roy:basu} respectively.

With regards to NC Landau problem, the key that usually being highlighted is to be able to show some energy correction from the Landau energy spectrum~\cite{Gam:Men, duka, KaHuaYan}. The computation can be retrieved from either only taking the spatial-spatial being noncommute or both spatial-momenta do not commute. The purpose of this paper is however to show that the energy correction that arise in the noncommutative Landau problems can just be retrieved directly after going thorough a simple algebraic manipulation problem.

The notion of noncommutative quantum mechanics or NCQM for short, can be understood by the extra feature added to the original quantum mechanics canonical commutation relation (CCR) between the position and momentum coordinates. Therefore, NCMQ can only be realized in two- and higher-dimensional system. The (non-unique) algebra which described noncommutative model is given by
\be\label{e:ncom}
[x_{i},x_{j}] = \ic\theta_{ij}; \quad [p_{i},p_{j}] = \ic\zeta_{ij}; \quad [x_{i},p_{j}]= \ic\hbar\delta_{ij}.
\ee
In literature, algebra (\ref{e:ncom}) is called the deformed Heisenberg algebra~\cite{KaHuaYan}. For $i,j=1,2$, $\delta_{ij}$ is the standard Kronecker delta while both $\theta_{ij}$ and  $\zeta_{ij}$ are known as antisymmetric noncommutative parameter which measure the noncommutativity between position, $x_i$ and momenta $p_i$ coordinates. This is of course analogous to the Planck constant that measure the noncommutativity between the position and momentum coordinate in the CCR. 

This paper is organized as follows. In Section 2, the algebraic method of the Landau problem following~\cite{ali} is highlighted for which the formulation will be important to construct results in Section 3. In Section 3, by using a proper Bopp shift we present the generalized Hamiltonian associated to the Landau problem. Conclusions are given in the last section.

\section{Landau problem from algebraic point of view}

For the Landau quantization problem the Hamiltonian (in some conveniently chosen units) can be written as 
\be
H_\text{elec} = \frac 12 (\vec p - \vec A )^2 = \frac 12 \left(p_x + \frac y2 \right)^2 +
      \frac 12 \left(p_y - \frac x2 \right)^2,
      \label{elec-ham}
\ee
where $\vec{A}$ is the vector potential together with the position $x,y$ and momentum $p_{x},p_{y}$ operators. As properly analyzed by~\cite{ali}, using the von Neumann algebraic property for the Landau levels problem, their corresponding Hamiltonian can be divided into two types that is described by an upward and downward  (direction) of the uniform magnetic field perpendicular to the $z$-axis. This result will basically be our starting point. 
 
It is shown that, since there are two Hamiltonians i.e. $H^{{\uparrow}}_{\text{elec}}$ and $H^{{\downarrow}}_{\text{elec}}$ which describe the electron on the plane, one can rewrite Eq. (\ref{elec-ham}) on the Hilbert space $L^2 (\mathbb R^2 , dxdy )$ such that
\begin{align*}
 p_x + \frac y2 \longrightarrow Q_1 &= -\ic\frac \partial{\partial x } + \frac y2 \; ,\\
 p_y - \frac x2 \longrightarrow P_1 &= -\ic\frac \partial{\partial y } - \frac x2 \; ,
\end{align*}
together with the following set
\begin{align*}
 Q_2 = -\ic\frac \partial{\partial y } + \frac x2; \qquad
 P_2 = -\ic\frac \partial{\partial x } - \frac y2 \; .
\end{align*}

 With this replacement, it easily verified that $[Q_i, P_j ] = \ic\text{I}\delta_{ij} $ for $i,j=1,2$.  For brevity, this will enable us to write the Hamiltonian as 
\be\label{lanElec}
  H^{\uparrow\downarrow}_\text{elec} = \frac 12 \left( P_{1,2}^2 + Q_{1,2}^2\right)\; .
\ee
where $\{Q_{1},P_{1} \}, \{ Q_{2}, P_{2}\}$ described $H^{\uparrow}_{\text{elec}}$ and $H^{\downarrow}_{\text{elec}}$ respectively.
Together, the two sets of operators $\{ Q_1 , P_1\}$ and $\{ Q_2 , P_2\}$ are mutually commute namely
$$
  [Q_2 , Q_1 ] = [P_2 , Q_1 ] = [Q_2 , P_1 ] = [P_2 , P_1 ] = 0 \; .
$$
By an operator method, we defined the operators,
$$
 A_1  =   \frac 1{\sqrt{2}} (\ic Q_1 - P_1),   \quad A_2   =   \frac 1{\sqrt{2}} (\ic Q_2 + P_2)
$$
and their adjoints, which then satisfy the only survive commutation relations,
$$
 [A_i , A^\dag_i ] = 1 , \qquad\text{for}\quad i =1,2.
$$
The eigenstates of the Hamiltonian are thus given by
$$
\Psi_{\ell ,  n}:=\frac{1}{\sqrt{n!\ell!}}\left(A_1^\dag\right)^n\left(A_2^\dag\right)^\ell\Psi_{0 0},
$$
where $\ell, n =0,1,2,\ldots$ and $\Psi_{0 0}$ is the ground state for which $A_1\Psi_{00}=A_2\Psi_{00}=0$.

The solutions also give rise to the Hermite polynomials for the case of two-variables. In addition, the  eigenvalues of this Hamiltonian, the so-called {\em Landau levels\/,}  $E_\ell = (\ell + \frac 12 ), \;\ell =0,1,2, \ldots \infty$, with  each level being  infinitely degenerate~\cite{ali} .

\section{Noncommutative Landau levels problem}

We next discuss the problem of the electron in a constant magnetic field, for which we had earlier obtained the energy levels using the Hamiltonian (\ref{elec-ham}) but under the noncommutative configuration (phase) space. We now rewrite this Hamiltonian, by replacing the standard position operators by the non-commutative ones. In literature~\cite{duka, diao} the noncommutative coordinates are deformed or have been shifted by the so-called ``Bopp shift'' such that
\begin{equation}\label{e:ncnew}
\widehat{Q}_{1} = q_{x} - \frac{\Theta}{2}p_{y}, \quad \widehat{Q}_{2} = q_{y} +\frac{\Theta}{2}p_{x}, \quad \widehat{P}_{1}=p_{x}, \;\; \widehat{P}_{2}=p_{y}.
\end{equation}
The only survive commutators are thus
\[
[\widehat{Q}_{1}, \widehat{Q}_{2}] = \Theta, \quad [\widehat{Q}_{i}, \widehat{P}_{j}]= \ic\delta_{ij}.
\]
We denote the noncommutative operators with a ``hat'' notation to differentiate with the commutative coordinates and now we use ${q_{x},q_{y}}$ to replace ${x,y}$.  For only taking spatial being noncommute, we arrive at
\be\label{e:nclandau}
H^{nc}_{elec} = \frac{1}{2}(\vec{\widehat{P}} - \vec{\widehat{A}})^{2}= \frac{1}{2}\Big(\widehat{P}_{1} + \frac{\widehat{Q}_{2}}{2} \Big)^{2} + \frac{1}{2}\Big( \widehat{P}_{2}-\frac{\widehat{Q}_{1}}{2}\Big)^{2}.
\ee
Substituting (\ref{e:ncnew}) in (\ref{e:nclandau}) yield
\begin{align*}\label{e:ncHam}
H^{nc}_{elec} &= \frac{1}{2}\Big(p_{x}+ \frac{q_{y}}{2} + \frac{\Theta}{4}p_{x}\Big)^{2} + \frac{1}{2}\Big(p_{y}- \frac{q_{x}}{2} + \frac{\Theta}{4}p_{y}\Big)^{2}, \\
\quad&=\frac{1}{2} \Big[(1+\frac{\Theta}{4})p_{x} + \frac{q_{y}}{2} \Big]^{2}+ \frac{1}{2}\Big[ (1+\frac{\Theta}{4})p_{y} - \frac{q_{x}}{2} \Big]^2.
\end{align*}
This denote the noncommutative Hamiltonian or simply generalized Hamiltonian. 
We further simplify by letting $\gamma =(1+\frac{\Theta}{4})$ and introduce the following operators
\be\label{e:wideop}
\widetilde{Q}_{1} = \sqrt{\gamma}p_{x} +\frac{1}{\sqrt{\gamma}} \frac{q_{y}}{2}, \quad \widetilde{P}_{1} =  \sqrt{\gamma}p_{y} -\frac{1}{\sqrt{\gamma}} \frac{q_{x}}{2},
\ee
then
$$
[\widetilde{Q}_{1}, \widetilde{P}_{1}] = -\frac{1}{2} [p_{x},q_{x}] + \frac{1}{2}[q_{y},p_{y}] = \ic.
$$
\noindent Finally we can rewrite (\ref{e:nclandau}) to become
\begin{align*}
H^{nc}_{elec}& = \frac{1}{2} \gamma \Big[ \Big(\sqrt{\gamma}p_{x} +\frac{1}{\sqrt{\gamma}} \frac{q_{y}}{2}\Big)^{2} + \Big(\sqrt{\gamma}p_{y} -\frac{1}{\sqrt{\gamma}} \frac{q_{x}}{2} \Big)^{2}\Big]\\
&= \frac{1}{2}(1+\frac {\Theta}4)(\widetilde{P}_{1}^{2}+\widetilde{Q}_{1}^{2}).
\end{align*}
Observing this form, we conclude that the effect of non-commutativity is to shift all the energy levels by the amount $\frac{\Theta}{8}$:
\be
\Delta E = \frac{\Theta}{8}. 
\ee

However, if one would introduce the ladder operator associated to (\ref{e:wideop}), then
\be
\widetilde{A}_{1}:= \dfrac{1}{\sqrt{2}}(\ic\widetilde{Q}_{1}-\widetilde{P}_{1}); \quad \widetilde{A}_{1}^{\dag}:= \dfrac{1}{\sqrt{2}}(-\ic\widetilde{Q}_{1}-\widetilde{P}_{1}),
\ee
such that $[\widetilde{A}_{1}, \widetilde{A}_{1}^{\dag}]=1$ together with its second set. This form is equivalent to the ordinary ladder operators for the commutative case. Unlike results appeared in~\cite{kangli05}, which observed a deformed ladder operator such that one will obtained $[\widetilde{A}_{i}, \widetilde{A}^{\dag}_{j}]=\delta_{ij}+\ic\theta_{ij}$.

As can be easily seen, introducing an additional non-commutativity, where the two observables of momentum also do not commute, the energy would be shifted by an additional amount that is in terms of the noncommutative parameter $\Theta$. This is a generalized form of the shifted energy associated to NC phase space which explicitly take the form of $\frac{Be\theta}{8c\hbar}$ according to~\cite{min}.

\section{Conclusion}

In this paper we study the Landau problem in noncommutative configuration space and phase space. The consideration of the NC space and NC phase space produces additional terms. With a proper defined Bopp shift, the correction to the usual Landau energy levels is obtained. It is shown that in the NC phase space the energy of the Landau levels is shifted by the amount of $\frac{\Theta}{8}$ and it is directly obtained from the generalized Hamiltonian that describe the effect of NC Landau levels problem. Our result is in fact different from the one discussed in~\cite{Gam:Men, duka}.


\end{document}